\documentclass[runningheads]{llncs}
\usepackage[T1]{fontenc}
\usepackage{graphicx}
\usepackage{tcolorbox}
\usepackage{multirow}
\usepackage{cite}
\usepackage{url}
\usepackage{hyperref}
\usepackage{pgfplots}
\usepackage{textcomp}
\usepackage{placeins}
\usepackage{fontawesome}
\usepackage{comment}
\usepackage{tabularx}
\usepackage{enumitem}
\usepackage{booktabs}
\usepackage{orcidlink}
\begin{document}
\title{On the Limitations of Combining Sentiment Analysis Tools in a Cross-Platform Setting}
\author{Martin Obaidi\inst{1}\orcidlink{0000-0001-9217-3934}\textsuperscript{\faEnvelopeO} \and
Henrik Holm\inst{1} \and
Kurt Schneider\inst{1}\orcidlink{0000-0002-7456-8323}\and
Jil Klünder\inst{1}\orcidlink{0000-0001-7674-2930}}
\authorrunning{Obaidi et al.}
\institute{Leibniz University Hannover, Software Engineering Group, Welfengarten 1, 30167 Hannover, Germany \\
\email{martin.obaidi / jil.kluender / kurt.schneider @inf.uni-hannover.de}
\email{hello@henrikholm.de}
}
\maketitle              % typeset the header of the contribution
\begin{abstract}
A positive working climate is essential in modern software development. It enhances productivity since a satisfied developer tends to deliver better results. Sentiment analysis tools are a means to analyze and classify textual communication between developers according to the polarity of the statements. Most of these tools deliver promising results when used with test data from the domain they are developed for (e.g., GitHub). But the tools' outcomes lack reliability when used in a different domain (e.g., Stack Overflow).
One possible way to mitigate this problem is to combine different tools trained in different domains. 
In this paper, we analyze a combination of three sentiment analysis tools in a voting classifier according to their reliability and performance. The tools are trained and evaluated using five already existing polarity data sets (e.g. from GitHub).
The results indicate that this kind of combination of tools is a good choice in the within-platform setting. However, a majority vote does not necessarily lead to better results when applying in cross-platform domains. In most cases, the best individual tool in the ensemble is preferable. This is mainly due to the often large difference in performance of the individual tools, even on the same data set. However, this may also be due to the different annotated data sets.

\end{abstract}
%
%
%

%todo
% SMS zitieren
% prelabeled data sets zitieren

\section{Introduction}
The application of sentiment analysis in software engineering (SE) has many purposes and facets ranging from identifying the actual mood in a team to extracting information from app-reviews \cite{obaidiSentiSMS22, linSentiSLR22}. These tools usually analyze statements for the pre-dominant sentiment and classify them according to their polarity (\textit{positive}, \textit{negative} and \textit{neutral}). Sentiment analysis tools are frequently applied to investigate the social component of a developer team (e.g., \cite{10.1002/smr.1673,SCHNEIDER201859}. 

These sentiment analysis tools are most often trained on data emerging from frequently used data bases such as Stack Overflow or GitHub~\cite{obaidiSentiSMS22, linSentiSLR22}. 
Within these different SE specific domains (e.g., GitHub), tools like Senti4SD \cite{Calefato.2018} or RoBERTa \cite{liu2019roberta} achieve high accuracies \cite{9240704,noviellicross20}. 

For a tool to be widely used in practice, it is important that it performs sufficiently well among different SE specific domains, and not just within one. Otherwise, it would be necessary to develop or train a separate tool for each domain, complicating the use in practice and reducing the number of application scenarios to a minimum. 

However, several papers indicate that SE specific sentiment analysis tools perform worse in domains in which they were not trained (e.g., \cite{noviellicross20,9240704,10.1145/3180155.3180195}), meaning that a tool trained and tested in one SE specific domain (e.g., GitHub data) performs worse in another domain (e.g., JIRA data). Cabrera-Diego et al. \cite{CABRERADIEGO2020105633} identified that tools trained with JIRA did not perform well on Stack Overflow. 
Novielli et al. \cite{noviellicross20} investigated the performance of pre-trained sentiment analysis tools in different, unknown domains, in a so called cross-platform setting. They overall observed a poor cross-platform performance of the tools. For example, they observed a 26\% difference in the macro average F1 score in a cross-platform setting with the tool Senti4SD \cite{Calefato.2018}.

One possible solution suggested by Obaidi et al. \cite{obaidiSentiSMS22} and Zhang et al. \cite{9240704} is to combine different tools by using a majority vote. The basic idea is that by combining different tools, a majority vote may be able to classify correctly, if a well-performing tool alone misclassifies. Moreover, such an ensemble combining different ``domain expert knowledge'' would allow the use in all domains instead of needing a separate tool for each domain -- if the approach works.

In this paper, we analyze how such a combination of different tools performs compared to single tools and compared to different settings of training and test data. That is, we present combinations of three different sentiment analysis tools in a voting classifier (VC). In a first step of our approach, we combine three sentiment analysis tools and train them with three data sets. In this scenario, one tool at a time is an ``expert'' for a SE specific domain (as the tools are specifically designed to be trained with one of the data sets). We first investigated whether the tools achieve a better accuracy on the within-platform when they are combined. Then, based on a quantitative approach, we tried several combinations of pre-trained tools to investigate their accuracy in a cross-platform setting in two different experiments.  

\textit{Outline.} The rest of the paper is structured as follows: In Section \ref{sec:background}, we present related work and background details. The concept of the voting classifier and its application in the study is introduced in Section \ref{sec:research}. Section \ref{sec:results} summarizes the results that are discussed in Section \ref{sec:discussion}, before concluding the paper in Section \ref{sec:conclusion}.

\section{Background and Related Work}
\label{sec:background}

In this section, we present related work on sentiment analysis tools in general, voting classifiers and on SE specific data sets used for sentiment analysis.

\subsection{Sentiment Analysis}

For the field of software engineering, several sentiment analysis tools have been developed or evaluated.

Calefato et al. \cite{Calefato.2018} developed the tool Senti4SD and thus enabled training and classification of models specific to the domain of SE. To do this, they used a specialized word lexicon in combination with a support-vector machine. With this approach, they were able to classify an input document in one of the three polarities \textit{positive, negative, and neutral}.

Zhang et al. \cite{9240704} compared the performance of different pre-trained neural network models with those tools using more classical machine learning approaches (but without training and thus without adapting them for each domain). These included four neural network models like RoBERTa \cite{liu2019roberta}, and five other existing tools (e.g., Senti4SD \cite{Calefato.2018}). They found that the performance of these tools changes depending on the test data set \cite{9240704}. They observed that the RoBERTa model \cite{liu2019roberta} most often had the highest scores on average among the pre-trained transformer models.

Novielli et al. \cite{noviellicross20} investigated in their cross-platform study to what degree tools that have been trained in one domain perform in another unknown domain. They used three data sets and four tools and concluded that supervised tools perform significantly worse on unknown domains and that in these cases a lexicon-based tool performs better across all domains.

In their replication study, Novielli et al. \cite{Novielli.replication2021} explained some sentiment analysis tools (e.g. Senti4SD \cite{Calefato.2018}) in great detail and described the underlying data. They also investigate the agreement between sentiment analysis tools with each other and with manual annotations with a gold standard of 600 documents. Based on their results, they suggest platform-specific tuning or retraining for sentiment analysis tools \cite{Novielli.replication2021}.

All these mentioned tools perform well within one domain, but significantly worse in cross-platform domains \cite{noviellicross20,9240704}. One possibility to counteract this is to use a combination of several tools. To the best of our knowledge, this approach has not yet been used for a cross-platform settings.

\subsection{Voting Classifier}

To the best of your knowledge, the concept of majority voting has been applied in the context of sentiment analysis in SE in only three papers.

Herrmann and Klünder \cite{9582407} applied a voting classifier (SEnti-Analyzer) consisting of three machine learning methods for German texts especially for meetings \cite{9582407}. They used three different machine learning algorithms and an evolutionary algorithm.

Uddin et al. \cite{uddinVoting2021} used different sentiment analysis tools in a majority voting ensemble. They combined tools such as Senti4SD \cite{Calefato.2018}, or RoBERTa \cite{liu2019roberta} in an ensemble and investigated whether majority voting can improve the performance of this ensemble compared to the best individual tool. In doing so, they combined several data sets into one large benchmark data set. Overall, they conclude that while ensembles can outperform the best individual tool in some cases, they cannot outperform it overall. 

However, neither paper specifically examined how tools trained in several different domains perform together in a cross-platform setting.

\subsection{SE Data Sets for Sentiment Analysis}
\label{sec:dataset}

Several papers highlight the need of domain adaptation to the field of SE (e,g., \cite{Calefato.2018}), leading to some SE specific data sets. Recent SMSs and SLRs about sentiment analysis in SE show an overview of the data sets used \cite{obaidiSentiSMS22,linSentiSLR22}

Novielli et al. \cite{Novielli.2018b} collected 4,800 questions asked on the question-and-answer site Stack Overflow and assigned emotions to each sentence of the collected communication. Afterwards, they labeled these sentences based on these emotions with three polarities \textit{positive}, \textit{negative} and \textit{neutral}. This labeling process was done by a majority decision of three raters.

Another gold standard data set crawled from GitHub was developed by Novielli et al. \cite{novielligold.2020}. This data set contains over 7,000 sentences. Similar to the Stack Overflow data set \cite{Novielli.2018b}, they first assigned emotions to each sentence and labeled polarities based on these emotions.

Ortu et al. \cite{Ortu.2016} developed a data set consisting of about 6,000 comments crawled from four public JIRA projects. They assigned each statement an emotion label based on the Parrott emotion model \cite{parrott2001emotions}.

Lin et al. \cite{10.1145/3180155.3180195} collected 1,500 discussions on Stack Overflow tagged with Java. Five authors labeled the data supported by a web application they built. In their paper no emotion model or guidelines for labeling were mentioned.

Uddin et al. \cite{8643972} developed the API data set. It consists of 4,522 sentences from 1,338 Stack Overflow posts regarding API aspects. The authors did not follow an emotion model or any guidelines, but in their coding guide, an example sentence was mentioned for each polarity with focus on the opinion expressed in the text. 

The APP reviews data set, labeled by Lin et al. \cite{10.1145/3180155.3180195}, consists of 341 mobile app reviews. No emotion model or guidelines for labeling are mentioned.

\section{Study Design}
\label{sec:research}

In this section, we present our research questions, our voting classifier approach and the used data sets. Afterwards, we describe our training methods, the used metrics for evaluation and for the quantitative analysis. 

\subsection{Research Questions}
\label{sec:researchquestions}

We chose five different data sets and developed a tool which combines three different sentiment analysis tools in a voting classifier. For this investigation, we pose the following research questions:
\\

\noindent \textbf{RQ1:} \textit{How does the classification accuracy of an ensemble of sentiment analysis tools vary in comparison to the individual tools trained and tested within the same domain?}

This allows us to evaluate whether a voting classifier in an already known domain offers any additional benefit in terms of performance. Based on the outcome of this research question, it would be conceivable to take an ensemble of individual tools for a certain domain in order to achieve a better classification accuracy.
\\

\noindent \textbf{RQ2:} \textit{How does the classification accuracy of the voting classifier vary in a cross-platform setting?}

To explore this research question in more detail, we split it into two parts:

\noindent \textbf{RQ2.1:} \textit{Do different tools, pre-trained on different domains, perform better in a cross-platform setting in an ensemble than the best individual tool?}

In this first experiment, we consider all possible combinations of tools and pre-trained three data sets, testing them on two other data sets.

\noindent \textbf{RQ2.2:} \textit{Do the best tools for different domains in an ensemble perform better in a cross-platform setting than the best individual tool?}

In the second experiment, we determine the best tool for each data set and then test different combinations of these tools, in a cross-platform setting. It can therefore happen that a tool occurs several times in an ensemble.

We call the analysis for the research questions RQ2.1 and RQ2.2 "experiment" so that we can simply refer to it in the further course of our study.

\subsection{Selection of Sentiment Analysis Tools}
\label{tools}
For our voting classifier, we selected three tools with different machine learning approaches which we described in Section \ref{sec:background}. Therefore, it is likely that they complement each other well regarding the strengths and weaknesses of these methods.

\begin{itemize}
    \item \textit{Senti4SD} \cite{Calefato.2018}, a machine learning based tool which uses support-vector machine and heuristics. It was originally developed for the analysis of Stack Overflow data.
    \item  \textit{RoBERTa} \cite{liu2019roberta}, which is a BERT-based model, using a neural network. RoBERTa was trained with several data sets consisting of, among others, Wikipedia articles, books and news.
    \item \textit{SEnti-Analyzer} \cite{9582407}, which implements different machine learning algorithms (random forest, naive Bayes, support-vector machine) and an evolutionary algorithm. This tool was developed specifically for software project meetings and trained with student communication data from software projects.
\end{itemize}

Senti4SD \cite{Calefato.2018} and SEnti-Analyzer \cite{9582407} were trained and fine tuned with their default settings. The settings for RoBERTa \cite{liu2019roberta} were adopted from the previous work of Zhang et al. \cite{9240704} as they yielded the best accuracies.

Each of the tools is combined in a majority voting. If there was a disagreement between all of them (and therefore no majority voting possible), the output label is set randomly.

\subsection{Data Sets}

We used a total of five different data sets for training and testing. The data sets are described in Table \ref{tab:datasets}. \#Docs stands for the number of statements in the respective data set.

\setlength{\tabcolsep}{7pt}
\begin{table}[htbp!]

\caption{Overview of the used data sets}
\label{tab:datasets}
%\begin{tabular}{p{3.2cm}p{1.4cm}lp{2.4cm}llll}
\begin{tabular}{@{}lllll@{}}
\toprule
Data set            & \#Docs        & \#Positive (\%) & \#Neutral (\%)  & \#Negative (\%) \\ \midrule
API                 & 4522          & 890 (19.7\%)    & 3136 (69.3\%)   & 496 (11\%)       \\
APP                 & 341           & 186 (54.5 \%)   & 25 (7.3\%)      & 130 (38.1\%)     \\
GitHub (G)          & 7122          & 2013 (28.3\%)   & 3022 (42.4\%)   & 2087 (29.3\%)    \\
JIRA (J)            & 3974          & 290 (7\%)       & 3058 (77\%)     & 626 (16\%)       \\
Stack Overflow (SO) & 4423          & 1527 (35\%)     & 1694 (38\%)     & 1202 (27\%)      \\ \bottomrule
\end{tabular}
\end{table}

Unlike the other data sets, the JIRA statements \cite{Ortu.2016} have emotions as a label. We took this data set and assigned polarities for each document similar to Calefato et al. \cite{Calefato.2018} corresponding to emotions. Since there were multiple duplicates in the data set, we ended up with a total of 3974 statements.

\subsection{Training}
We trained each tool with 5-fold cross-validation. The data sets were shuffled and then split into five stratified folds. These stratified folds were balanced just like the data set. By using this kind of validation, we have decided against balancing the data sets, because it seems to be "realistic" that there is a lot of neutral in many data sets, which is the case for most data sets we use. The tools themselves implemented certain procedures to deal with unbalanced data as well. 

\subsection{Evaluation Metrics}
\label{subsec:evaluation}

When evaluating a model, accuracy is considered as the first relevant measurement value, since it represents the overall classification accuracy of a model \cite{infretbasics}. Therefore, we present the accuracy in our results. However, for the choice of the best tool or model, we also consider the macro average F1 value ($F1_{macro}$), because it considers both the precision and the recall of all classes. 

\subsection{Interrater Agreement}
\label{sub:interrater}
To measure the extent to which the individual classification tools used within the voting classifier agree in their decisions, we calculated \textit{Fleiss' Kappa} ($\kappa$) \cite{fleiss1971measuring} as agreement value. Fleiss' kappa measures the interrater reliability between more than two raters \cite{fleiss1971measuring}. We classify the kappa value based on Landis and Koch \cite{landis1977measurement}.

\subsection{Combination of the Tools}
\label{sec:combination}
Table \ref{tab:vcconfigs1} shows all the combinations of tools and training data sets we used for the evaluation of RQ1. Here we test the voting classifier in a within domain setting.

% für eventuell RQ4 auch das hier verwenden
\begin{table}[htbp!]
\small
\caption{Tested combinations of tools in a within domain setting with the same pre-trained data set}
\label{tab:vcconfigs1}
\begin{tabular}{lllll}
\toprule
ID & Senti4SD       & RoBERTa        & SEnti-Analyzer            \\ \midrule
1  & GitHub         & GitHub         & GitHub         \\
2  & JIRA           & JIRA           & JIRA           \\
3  & Stack Overflow & Stack Overflow & Stack Overflow \\
4  & API            & API            & API            \\
5  & APP            & APP            & APP            \\ \bottomrule
\end{tabular}
\end{table}
We assigned an ID to each of the different combinations of tools so that we can reference them in the results later. 

For the research questions RQ2.1 and RQ2.2, we opted for a quantitative approach to ensure the widest possible coverage of combinations. The different combinations for the first experiment (RQ2.1) are presented in Table \ref{tab:vcconfigs2}.
\begin{table}[htbp!]
\small
\caption{First experiment for testing combinations of tools with the respective cross-platform domains for RQ2.1}
\label{tab:vcconfigs2}
\begin{tabular}{lllll}
\toprule
ID & Senti4SD       & RoBERTa        & SEnti-Analyzer            \\ \midrule
6  & GitHub         & Stack Overflow & JIRA           \\
7  & GitHub         & JIRA           & Stack Overflow \\
8  & Stack Overflow & GitHub         & JIRA           \\
9  & Stack Overflow & JIRA           & GitHub         \\
10  & JIRA           & GitHub         & Stack Overflow \\
11  & JIRA           & Stack Overflow & GitHub         \\
\bottomrule
\end{tabular}
\end{table}

Here we selected three different tools and then permuted the respective domains within these tools. 

In the second experiment (RQ2.2), we chose a deeper analysis and selected the best performing tool per domain, regardless of whether it was already selected for another domain as well. Table \ref{tab:vcconfigs3} presents the list of combinations. Because in all evaluations RoBERTa was the best performing tool, we listed the three tools as ``RoBERTa1'', ``RoBERTa2'' and ``RoBERTa3''.

\begin{table}[htbp!]
\small
\caption{Second experiment for testing combinations of tools with the respective cross-platform domains for RQ2.2}
\label{tab:vcconfigs3}
\begin{tabular}{lllll}
\toprule
ID & RoBERTa1      & RoBERTa2     & RoBERTa3            \\ \midrule
12  & GitHub         & JIRA         & Stack Overflow        \\
13  & GitHub         & API           & JIRA           \\
14  & GitHub         & Stack Overflow & API \\
15  & Stack Overflow & JIRA            & API            \\
\bottomrule
\end{tabular}
\end{table}

We marked each usage of a combination of tools with a number after the ID, separated by a dot. For example, if the combination with ID 2 was used 3 times, we wrote ID 2.1, 2.2 and 2.3 in the tables and in the text. This ensures that we can definitively identify each run.

\section{Results}
\label{sec:results}
We conducted the data analysis as described in Section \ref{sec:research}. In the following, we present the results for the different analysis steps.

\subsection{Within-domain classification}

The results of the analysis related to RQ1 can be found in Table \ref{tab:samedomain}. As mentioned before, disagreement means that all three tools have assigned a different label (see \ref{tools}).

\setlength{\tabcolsep}{3.5pt}
%\parbox{1.5cm}{SEnti-\\ Analyzer}
\begin{table}[htbp!]
\small
\caption{Classification accuracy within the same domain}
\begin{tabular}{@{}llllllll@{}}
\toprule
ID      & \#Data  &    VC             & Senti4SD    & RoBERTa        & SEnti-Analyzer  & Disagreement   & $\kappa$ \\ \midrule
1.1 (G)   &  1426   &   \textbf{0.93}   & 0.92        & 0.92           & 0.88                                 & 0.8\%   & 0.83 \\
2.1 (J)    &  796   &   \textbf{0.89}   & 0.85        & 0.87           & 0.88                                 & 0.1\%  & 0.68  \\
3.1 (SO)  &  885    &   \textbf{0.90}   & 0.87        & \textbf{0.90}  & 0.86                                 & 1.2\%   & 0.80  \\
4.1 (API) &  904    &   \textbf{0.91}   & 0.87       & 0.89            & 0.86                                 & 0.9\%   & 0.68  \\
5.1 (APP) &   69    &   \textbf{0.88}   & 0.87       & 0.87            & 0.86                                 & 0.0\%    & 0.84  \\
 \bottomrule
\end{tabular}
\label{tab:samedomain}
\end{table}

For four out of five data sets, the VC achieved an accuracy 1-2\% higher than the best individual tool in its ensemble (ID 1.1, 2.1, 4.1 and 5.1). Only on Stack Overflow (ID 3.1), the accuracy of 90\% is identical between the voting classifier and the RoBERTa model.

The Fleiss' $\kappa$ value for these combinations range from 0.68 to 0.84, which shows that the tools have a substantial to almost perfect agreement. The relative amount of disagreements per document are under 1.2\%. \\

\subsection{Cross-platform Domains}
To answer RQ2.1 and RQ2.2 we conducted an experiment for each research question. 

\subsubsection{First Experiment}
\label{subsec:first}

The results of the evaluation on API \cite{8643972} and APP \cite{10.1145/3180155.3180195} are summarized in Table \ref{tab:rq21results}.

\setlength{\tabcolsep}{2.5pt}
\begin{table}[htbp!]
\small
\caption{Classification accuracies for RQ2.1. The best accuracy for each ID is highlighted in bold.}
\begin{tabular}{@{}lllllllll@{}}
\toprule
ID &Testset     &  \#Data & VC            & Senti4SD                &  RoBERTa            &  SEnti-Analyzer  & Disagreement & $\kappa$ \\ \midrule
6.1  & API & 4522   & \textbf{0.72} & \textbf{0.72} (G)       & \textbf{0.72} (SO)  & 0.70 (J)                & 0.4\%  & 0.33 \\ 
7.1  & API & 4522   & 0.73          & 0.72 (G)                & 0.70 (J)            & \textbf{0.74} (SO)                       & 0.9\%  & 0.33   \\ 
8.1  & API & 4522   & \textbf{0.72} & \textbf{0.72} (SO)      & 0.70 (G)            & 0.70 (J)                                 & 1.0\%  & 0.23\\ 
9.1  & API & 4522   &  \textbf{0.73} & 0.72 (SO)              & 0.70 (J)            & \textbf{0.73} (G)                        & 1.5\%  & 0.29\\ 
10.1 & API & 4522   & 0.73          & 0.70 (J)                & 0.70 (G)            & \textbf{0.74} (SO)                       & 1.0\%  & 0.28\\ 
11.1 & API & 4522   & 0.72          & 0.70 (J)                & 0.72 (SO)           & \textbf{0.73} (G)                        & 0.9\%  & 0.28\\ \midrule
6.2  & APP & 341    & 0.55             & 0.57 (G)   & \textbf{0.70} (SO)   & 0.25 (J)                               &  6.5\%    & 0.20     \\ 
7.2  & APP &  341    & 0.61             & 0.57 (G)   & 0.50 (J)             & \textbf{0.62} (SO)                     &  7.0\%    & 0.39   \\ 
8.2  & APP & 341    & 0.60             & 0.59 (SO)  & \textbf{0.77} (G)    & 0.25 (J)                               &  11.7\%      & 0.14   \\ 
9.2  & APP & 341    & \textbf{0.62}    & 0.59 (SO)  & 0.50 (J)             & 0.61 (G)                               &  4.7\%      & 0.42 \\ 
10.2 & APP & 341    & 0.52             & 0.35 (J)   & \textbf{0.77} (G)    & 0.62 (SO)                              &  12.0\%     & 0.24    \\ 
11.2 & APP & 341    & 0.61             & 0.35 (J)   & \textbf{0.70} (SO)   & 0.61 (G)                               &  5.9\%     & 0.31    \\ 
\bottomrule
\end{tabular}
\label{tab:rq21results}
\end{table}
Overall, the voting classifier is either worse or exactly as good as the best of the used individual tools within the ensemble. The highest accuracy achieved by the voting classifier for API is 73\% (IDs 9.1 and 10.1), and thus 1\% worse than the best individual tool. 
All individual tools trained with JIRA achieve notably worse accuracy values compared to the other two data sets for the APP domain. Only in one run (ID 9.2), the voting classifier is more accurate (62\%) than the three individual tools used in its ensemble. The RoBERTa model pre-trained with GitHub data achieves an accuracy of 77\% (ID 8.2 and 10.2) and, thus, has the highest overall accuracy, beating the best VC by 15\%.. 

The Fleiss' $\kappa$ values range from 0.14 to 0.42, which corresponds to slight to fair agreement. The random label assignment per data point (disagreement) is higher compared to RQ1. For ID 8.2 (11.7\%) and 10.2 (12\%.), they are even above 10\%, whereas the highest disagreement value in the API data set was 1.5\%, the lowest was even 0.4\%. Thus, the tools are less likely to agree on other domains which they have not been trained with.
\\

\subsubsection{Second Experiment}
\label{subsec:second}

In the second experiment, instead of three different tools, we selected the best tool for each domain. For this purpose, we calculated the average of accuracy and F1 score across all folds and chose the best ones. 
The combination IDs are presented in Table \ref{tab:vcconfigs3} of Section \ref{sec:combination} and the results are presented in Table \ref{tab:secondexperiment}.

\begin{table*}[htbp!]
\small
\caption{Classification accuracies of the tools for each SE domain on multiple data sets. The best accuracy for each ID is highlighted in bold.}
\begin{tabular}{@{}lllllllll@{}}
\toprule
ID & Testset     & \#Data  &  VC            & RoBERTa1           & RoBERTa2     & RoBERTa3              & Disagreement  & $\kappa $\\ \midrule
12.1 & API         & 4522    &  \textbf{0.71} & 0.70 (G)           & 0.70 (J)     & 0.70 (SO)             & 0.3\%   & 0.41     \\ 
12.2 & APP         & 341     &  0.70           & \textbf{0.77} (G)  & 0.50 (J)     & 0.65 (SO)             & 6.2\%   & 0.40     \\ 
13.1 & SO          & 4423    &  0.82          & \textbf{0.85} (G)  & 0.73 (API)   & 0.67 (J)              & 2.5\%   & 0.49      \\ 
13.2 & APP         & 341     &  0.73          & \textbf{0.77} (G)  & 0.67 (API)   & 0.50 (J)              & 7.9\%   & 0.39        \\ 
14.1 & J           & 3976   &  \textbf{0.79} & \textbf{0.79} (G)  & 0.76 (SO)    & 0.73 (API)            & 0.9\%   & 0.51    \\
14.1 & APP         & 341     &  0.76          & \textbf{0.77} (G)  & 0.65 (SO)    & 0.67 (API)            & 3.2\%   & 0.50    \\ 
15.1 & G           & 7122    &  0.75          & \textbf{0.79} (SO) & 0.67 (J)     & 0.63 (API)            & 3.8\%   & 0.42   \\
15.2 & APP         & 341     &  \textbf{0.67} & 0.65 (SO)          & 0.50 (J)     & \textbf{0.67} (API)   & 7.9\%   & 0.32   \\
\bottomrule
\end{tabular}
\label{tab:secondexperiment}
\end{table*}

In one out of eight runs, the voting classifier has the highest accuracy (ID 12.1), in two other cases (ID 14.1 and 15.2), it shares the first place with another tool. In the other five cases, it is second place.

Overall, the tools in experiment two performed the most robustly with a range of 0.67 to 0.82 (0.74 in average), compared to experiment one with a range of 0.55 to 0.73 (0.66). 
\\

\subsection{Comparison of Results from RQ1 and RQ2}

To study in more detail how the voting classifier performed among these two experiments, we built Table \ref{tab:comparisation} based on the results from RQ1 and RQ2. It shows the performance of the best voting classifier from RQ1 (VC RQ1). In contrast, it also shows the performance of the best individual tool from the first two experiments (Tool RQ2) as well as best the voting classifier (VC RQ2).

\setlength{\tabcolsep}{7pt}
\begin{table}[htbp!]
\small
\small
\caption{Performance of the best tools from RQ1 and RQ2 for each data set. The best accuracy for each data set is highlighted in bold.}
\begin{tabular}{@{}llll@{}}
\toprule
Data set    & VC RQ1            & Tool RQ2      & VC RQ2            \\ \midrule
GitHub (G)           & \textbf{0.93}     & 0.79 (SO)     & 0.75                            \\
JIRA (J)           &  \textbf{0.89}    & 0.79 (G)      & 0.79                          \\
Stack Overflow (SO)          &  \textbf{0.90}    & 0.85 (G)      & 0.82                           \\
API         &  \textbf{0.91}    & 0.74 (SO)     & 0.73                          \\
APP         &  \textbf{0.88}    & 0.77 (G)      & 0.76                          \\
 \bottomrule
\end{tabular}
\label{tab:comparisation}
\end{table}

The largest difference of accuracy between the voting classifier (0.91) and the best VC from RQ2 (0.73) is 15\% for API. The lowest is 8\% for Stack Overflow. Every single tool was better than the best voting classifier except for JIRA, where the best voting classifier and individual tool had 79\% accuracy. The deviations for the rest are between 1\% and 4\%. 
In three out of five cases, a tool pre-trained with GitHub performed best, for the other two it is Stack Overflow. No tool performed best pre-trained with JIRA or API.

\setlength{\tabcolsep}{7pt}
\begin{table}[htbp!]
\small
\small
\caption{Performance of the worst tools from RQ1 and RQ2 for each data set. The best accuracy for each data set is highlighted in bold.}
\begin{tabular}{@{}llll@{}}
\toprule
Data set    & VC RQ1            & Tool RQ2      & VC RQ2            \\ \midrule
GitHub (G)           & \textbf{0.93}     & 0.63 (API)     & 0.75                            \\
JIRA (J)           &  \textbf{0.89}    & 0.73 (API)      & 0.79                          \\
Stack Overflow (SO)  &  \textbf{0.90}    & 0.67 (J)      & 0.82                           \\
API         &  \textbf{0.91}    & 0.70 (G/J/SO)     & 0.71                          \\
APP         &  \textbf{0.88}    & 0.25 (J)      & 0.52                         \\
 \bottomrule
\end{tabular}
\label{tab:comparisationworse}
\end{table}

\section{Discussion}
\label{sec:discussion}

\subsection{Answering the Research Questions}
Based on the results and the findings of Section \ref{sec:results}, we answer the research questions as follows.
\\

\noindent \textbf{RQ1:} The results show that in most cases, we achieve higher classification accuracies using a voting classifier rather than using the individual tools in its ensemble. For four of the five data sets considered, the voting classifier achieves a 1-2\% higher accuracy than the best individual tool and 2-5\% higher than the worst tool.
\\

\noindent \textbf{RQ2:} The use of a voting classifier does not lead to an increase of accuracy in a cross-platform setting in most cases. For a voting classifier, it is most often better to choose the best performing tool for each SE domain (RQ2.2). However, in most cases, the best individual tool should be preferred. 

\subsection{Interpretation}

The basic idea of using a voting classifier can be useful, as our results showed with respect to RQ1. Hence, if the domain is already known and data sets exist for it, it is often useful to combine different, well-performing tools in one voting classifier, since the implementation of a voting classifier is quite simple.

However, based on the results of RQ2, there is not always an improvement by such a combination with different tools. Comparing RQ2.1 and RQ2.2, it can be concluded that the best method to achieve a good, robust performance of the voting classifier is to assemble the best tools for different data sets (RQ2.2).

It seems that the pre-trained data set plays a role in the performance of the single tool in an ensemble. This is not surprising at first glance. However, even though the API data contains statements from the Stack Overflow platform \cite{8643972}, tools pre-trained with API performs worse on the data set Stack Overflow \cite{Novielli.2018b} than tools pre-trained with the GitHub data set \cite{novielligold.2020} (e.g. ID 13.1). One reason could be the different labeling process and the subjectivity of it, as mentioned by many papers (e.g., \cite{herrmannSentiSurvey22, 8643972, 9240704, 10.1145/3180155.3180195}). 

Our results also indicate that labels from some data sets were subjectively labeled and may differ from labels from other data sets.
For example, the tools that were trained with JIRA in the first experiment all received the worst accuracies in both cross-platform domains. It is reasonable to assume that the JIRA data set is very different from the APP review data set. The JIRA data set is the oldest data set among the ones considered in our study (published with labels 2016 \cite{Ortu.2016}), so the labeling rigor back than might not have been the same. Another indication is the good performance of the tools, which were pre-trained by gold standard data sets. RoBERTa pre-trained with GitHub achieved the best accuracies in experiment two. For all four other cross-platform data sets (API, APP, Stack Overflow and JIRA), it was the best single tool. The GitHub data set differs from the other data sets as it is by far the largest data set and was first labeled by a guideline emotion model and then mapped to the polarities. This is also true for the Stack Overflow and JIRA data sets, but the JIRA data set is with 3974 statements smaller than GitHub (7122 statements) by almost 44\% and is very unbalanced in regard to the distribution of polarities(cf. Table \ref{tab:datasets}). 77\% of the statements are neutral, which is much more compared to GitHub (42.4\%). The Stack Overflow data set has almost 40\% less data (4423). Unsurprisingly, RoBERTa pre-trained with Stack Overflow is the second best performing tool for three out of four other cross-platform data sets.
Therefore, one reason for the good performance could be the large amount of data as well as the balanced distribution and the annotation process.
This supports the need for consistent, noise-free gold standard sets \cite{noviellicross20, N.Novielli.2018}. 
These two assumptions (much data and emotion guidelines annotation) is also supported by the fact that in the comparison of all tools and voting classifiers in RQ2, only those pre-trained with GitHub or Stack Overflow performed best (cf. Table \ref{tab:comparisation}).\\
On the other hand, while the JIRA data set is poorly balanced, it was labeled based on the Parrott emotion model \cite{parrott2001emotions}, but by other authors \cite{Ortu.2016} compared to GitHub or Stack Overflow. However, RoBERTa pre-trained with JIRA performed the worst compared to RoBERTa pre-trained with API in all cross-platform data sets, except for GitHub. Therefore, another explanation could be that people communicate differently on JIRA and during reviews, e.g., we observe different levels of politeness. Another reason could be the unbalanced distribution of the JIRA data set.\\
The performance of the voting classifier in the cross-platform setting shows the tendency that it rather does not matter whether we choose different tools and thus different machine learning approaches or not. Surprisingly, RQ2.1 has shown that different tools, pre-trained in the same domain, have larger accuracy differences among themselves in a cross-platform setting. As soon as one or two tools in an ensemble perform significantly worse compared to the others, the voting classifier will accordingly not perform better than the best individual tool, and vice versa. Since we do not necessarily know the accuracies of individual tools in a cross-platform setting (e.g., new unlabeled data), it is a matter of luck whether the voting classifier performs better or a single tool.
Therefore, we can conclude that if the domain is unknown, a voting classifier may well be an alternative to the dictionary-based tools that have been proposed so far for these cases \cite{noviellicross20}. But, our results do not show that a voting classifier can solve the problem of poor cross-platform performance, but rather show the limitations of such an approach in our setting. 

\subsection{Threats to Validity}
\label{sec:threats}
In the following, we present threats to validity according to our study. We categorize the threats according to Wohlin et al. \cite{Wohlin.2012} as internal, external, construct, and conclusion validity.

The 1-2\% better performance of the voting classifier compared to the best single tool observed in RQ1 is not high and thus could have occurred rather by coincidence (construct validity). However, the performance was 2-5\% higher than the worst tool in the ensemble and we performed the evaluation multiple times to minimize this threat.

For RQ2, we used two different cross-platform domains for testing. One of them is the API data set from Uddin et al. \cite{8643972}. This data is also from the platform Stack Overflow like the data set from Novielli et al. \cite{Novielli.2018b}, which we used for training (internal validity). However, the API data set only includes comments regarding API. Moreover, based on the results of RQ2, we found indications that a tool being pre-trained with Stack Overflow had not resulted in an advantage regarding the API data set. 

In the second experiment (RQ2.2), we only used RoBERTa. This makes the results of the voting classifier dependent on only one tool (construct validity). However, for each domain, based on the evaluation metrics, RoBERTa was the best performing tool.

In some circumstances it is possible that the results of this work are specific to the data sets used (construct validity). The API data set of Uddin et al. \cite{8643972} and the APP review data set by Lin et al. \cite{10.1145/3180155.3180195} were presumably labeled ad hoc. The label assignment may not be representative (construct validity). By using the three data sets from Novielli et al. \cite{novielligold.2020, Novielli.2018b} and Ortu et al. \cite{Ortu.2016}, we attempted to minimize this threat, as these three data sets were labeled based on the same emotion model.

\subsection{Future Work}
\label{sec:future}
Based on the results, interpretation, and to minimize the previously mentioned threats, we propose the following:

Evaluating other data sets is a possible approach to possibly improve the overall accuracy of the voting classifier. It would be beneficial to build more gold standard data sets for as many different SE specific domains as possible.
Besides, the data sets should be examined for subjective labels. It should be investigated whether these labels were assigned consistently and to what extent subjectivity plays a role. Is a meaningful assignment of polarities possible at all? Or is the degree of subjectivity too great? In addition, factors such as the context of sentences should be considered. Do developers pay more attention to the perceived tone (e.g. "@Mark Not you again...") or more to the content (e.g. "I don't like this phone")? Here it could be researched in the direction of linguistics and psychology, like Gachechiladze et al. \cite{10.1109/ICSE-NIER.2017.18} conducted. 
%A distinction should be made between label assignments based on emotion models and assignments made ad hoc. We should consider which of the two types is more practical and thus more meaningful to use

Furthermore, it may also be interesting to expand the number of tools within the voting classifier to allow for even greater domain coverage. 

Our goal in this work was to combine pre-trained experts, because it was already found by Novielli et al. \cite{noviellicross20} that dictionary tools often performed worse compared to pre-trained tools in the same domain. Nevertheless, it is possible that a mix of pre-trained tools as well as dictionary-based tools perform even better, since dictionary-based tools have the advantage of hardly performing very badly since they use a dictionary \cite{noviellicross20}. 

The statements of our results do not necessarily have to be limited to the field of sentiment analysis. An similar analysis on other classification areas (such as classification to bugs or feature requests) would also be interesting.

\section{Conclusion}
\label{sec:conclusion}

To successfully complete a software project, it's important that developers are satisfied. To be able to measure the sentiment in teams, sentiment analysis tools have been developed. To train these machine learning tools to be used in the SE domains, data from platforms such as GitHub or Stack Overflow were crawled and manually assigned to sentiments.

However, tools trained in one of these SE specific domains perform notably worse in another, cross-platform domain. We analyzed this issue in detail. We first investigated whether a voting classifier can improve the accuracy of the classification in the same domain. Our evaluation showed that a voting classifier could improve the accuracy in three cases by 1\% compared and in one case by 2\% to the best performing tool of its ensemble.

Afterwards, we have examined the behavior of the voting classifier in a cross-platform setting. For this purpose, we conducted two experiments. In both experiments, we observed that the use of the voting classifier did not lead to an increase in accuracy compared to the individual tools in most cases. There were constellations in which the voting classifier was better than the individual tools within this ensemble. However, the voting classifier was not able to prove itself in terms of accuracy in an overall comparison.  

When the three individual tools performed similarly well in the ensemble, the voting classifier was often the best performing. However, if one or two tools performed significantly worse than the rest in the ensemble, this had a corresponding influence on the voting classifier. Surprisingly, this deviation in performance was also observed and had a respective influence on the voting classifier when the individual tools were pre-trained with the same data. The influence of the data set chosen for pre-training has a more significant influence on the performance of the individual tools and thus on the voting classifier. 

\section*{Acknowledgment}
This research was funded by the Leibniz University Hannover as a Leibniz Young Investigator Grant (Project \textit{ComContA}, Project Number \textit{85430128}, 2020--2022).

%\subsubsection{Acknowledgements} Please place your acknowledgments at the end of the paper, preceded by an unnumbered run-in heading (i.e. 3rd-level heading).

\bibliographystyle{splncs04}
\bibliography{references}

\end{document}